\documentclass[
reprint,
nobibnotes,
bibnotes,
 amsmath,amssymb,
 aps,
floatfix
]{revtex4-2}
\usepackage{color}
\usepackage{mathrsfs}
\usepackage{url}
\usepackage[cmintegrals]{newtxmath}
\usepackage{graphicx}
\usepackage{dcolumn}
\usepackage{bm}
\usepackage[mathlines]{lineno}
\begin{document}

\preprint{APS/123-QED}

\title{Optical linewidth of soliton microcombs}

\author{Fuchuan Lei}
 \affiliation {Department of Microtechnology and Nanoscience, Chalmers University of Technology SE-41296 Gothenburg, Sweden}
\author{Zhichao Ye}%
 \affiliation {Department of Microtechnology and Nanoscience, Chalmers University of Technology SE-41296 Gothenburg, Sweden}
 \author{Óskar B. Helgason}%
 \affiliation {Department of Microtechnology and Nanoscience, Chalmers University of Technology SE-41296 Gothenburg, Sweden}
 \author{ Attila F{\"u}l{\"o}p}%
 \affiliation {Department of Microtechnology and Nanoscience, Chalmers University of Technology SE-41296 Gothenburg, Sweden}
  \author{Marcello Girardi}%
 \affiliation {Department of Microtechnology and Nanoscience, Chalmers University of Technology SE-41296 Gothenburg, Sweden}
 \author{Victor Torres-Company}%
  \email{torresv@chalmers.se}
 \affiliation {Department of Microtechnology and Nanoscience, Chalmers University of Technology SE-41296 Gothenburg, Sweden}

\date{\today}

\begin{abstract}
Soliton microcombs provide a versatile platform for realizing fundamental studies and technological applications. To be utilized as frequency rulers for precision metrology, soliton microcombs must display broadband phase coherence, a parameter characterized by the optical phase or frequency noise of the comb lines and their corresponding optical linewidths. Here, we analyse the optical phase-noise dynamics in soliton microcombs generated in silicon nitride high-Q microresonators and show that, because of the Raman self-frequency shift or dispersive-wave recoil, the Lorentzian linewidth of some of the comb lines can, surprisingly, be narrower than that of the pump laser. This work elucidates information about the physical limits in phase coherence of soliton microcombs and illustrates a new strategy for the generation of spectrally coherent light on chip.
\end{abstract}

\maketitle

The study of the laser’s linewidth started with the seminal works of Schawlow and Townes \cite{PhysRev.112.1940}, even before the invention of the laser. For an ideal laser oscillator, the frequency noise power spectral density (PSD) is a constant that determines the oscillator linewidth \cite{zhang2020general}. In practice, the PSD of a laser is far richer and complex \cite{di2010simple}. The laser linewidth derived from the flat region at high offset frequencies in the PSD is called intrinsic (or Lorentzian) as it arises from white frequency noise which cannot be effectively suppressed through a finite-bandwidth feedback control loop. The Lorentzian linewidth hence represents the ultimate performance in temporal coherence of an oscillator. Recent breakthroughs in silicon photonics have demonstrated integrated laser oscillators with a Lorentzian linewidth at the Hz (and below) level \cite{jin2021hertz,gundavarapu2019sub}. In optical coherent communications, the Lorentzian linewidth is a key metric, as it determines the instantaneous frequency and phase fluctuations on a short time scale that must be tracked with sufficient accuracy by the receiver for efficient distortion compensation \cite{pfau2009hardware}. 

A  frequency comb is a laser whose spectrum is composed of equidistant frequency components that are phase locked to a common frequency reference. The phase noise of the constituent optical lines sets a physical limit on the achievable time and frequency stability \cite{ludlow2015optical,diddams2020optical,fortier201920}. Significant efforts have been devoted to the systematic understanding of the  linewidth of mode-locked lasers  and frequency combs based on solid-state \cite{schmeissner2014spectral,liehl2019deterministic}, semiconductor \cite{takushima2004linewidth} and fiber lasers \cite{newbury2005theory,kim2016ultralow}. In 2007, a new type of frequency comb source (microcomb) was demonstrated \cite{del2007optical}. Microcombs harness the Kerr nonlinearity and large intensity buildup in a high-Q microresonator cavity. Low-noise coherent states can be attained through the generation of dissipative solitons \cite{leo2010temporal,herr2014temporal,kippenberg2018dissipative}. Unlike in conventional frequency combs based on mode-locked lasers, where the gain originates from stimulated emission in active gain media and the Lorentzian linewidth is partially dictated by spontaneous emission, the gain of soliton microcombs is based on resonantly enhanced continuous-wave-pumped parametric amplification, and the noise caused by spontaneous scattering is very weak. Another important difference is that in microcombs, the pump laser is coherently added to the comb spectrum, and therefore its noise is expected to be transferred equally to all comb lines. Indeed, earlier studies demonstrated that when the microcomb operates in a low-noise state, the comb lines inherit the linewidth of the pump \cite{del2011octave,liao2017dependence}, with lines further away degrading more due to  thermo-refractive noise (TRN) in the cavity \cite{drake2020thermal,nishimoto2020investigation}.

In this work, we present both theoretical and experimental studies of the Lorentzian linewidth of soliton microcombs. We reveal that the interplay between the pump's frequency noise and soliton dynamics results in a linewidth distribution among comb lines that is consistent with the elastic-tape model  \cite{telle2002kerr}, akin to what has been found previously with conventional mode-locked frequency combs \cite{takushima2004linewidth,liehl2019deterministic,kim2016ultralow}. An important difference in soliton microcombs is that the Raman self-frequency shift and dispersive-wave recoil couple the repetition rate of the soliton with the pump frequency  \cite{karpov2016raman,yi2017single}. This results in a subset of comb lines becoming resilient to the frequency noise of the pump laser, and an encompassing decrease in Lorentzian linewidth. 
Our work provides a comprehensive understanding of the optical phase noise dynamics in soliton microcombs. Furthermore, the mechanism for the reduction of the pump linewidth provides a new strategy to generate ultra-low-phase-noise coherent optical oscillators on chip.  \\

\noindent\textbf{Results}

\noindent\textbf{Elastic-tape model applied to soliton microcombs.}

\begin{figure*}[ht]
\centering
\includegraphics[width=\linewidth]{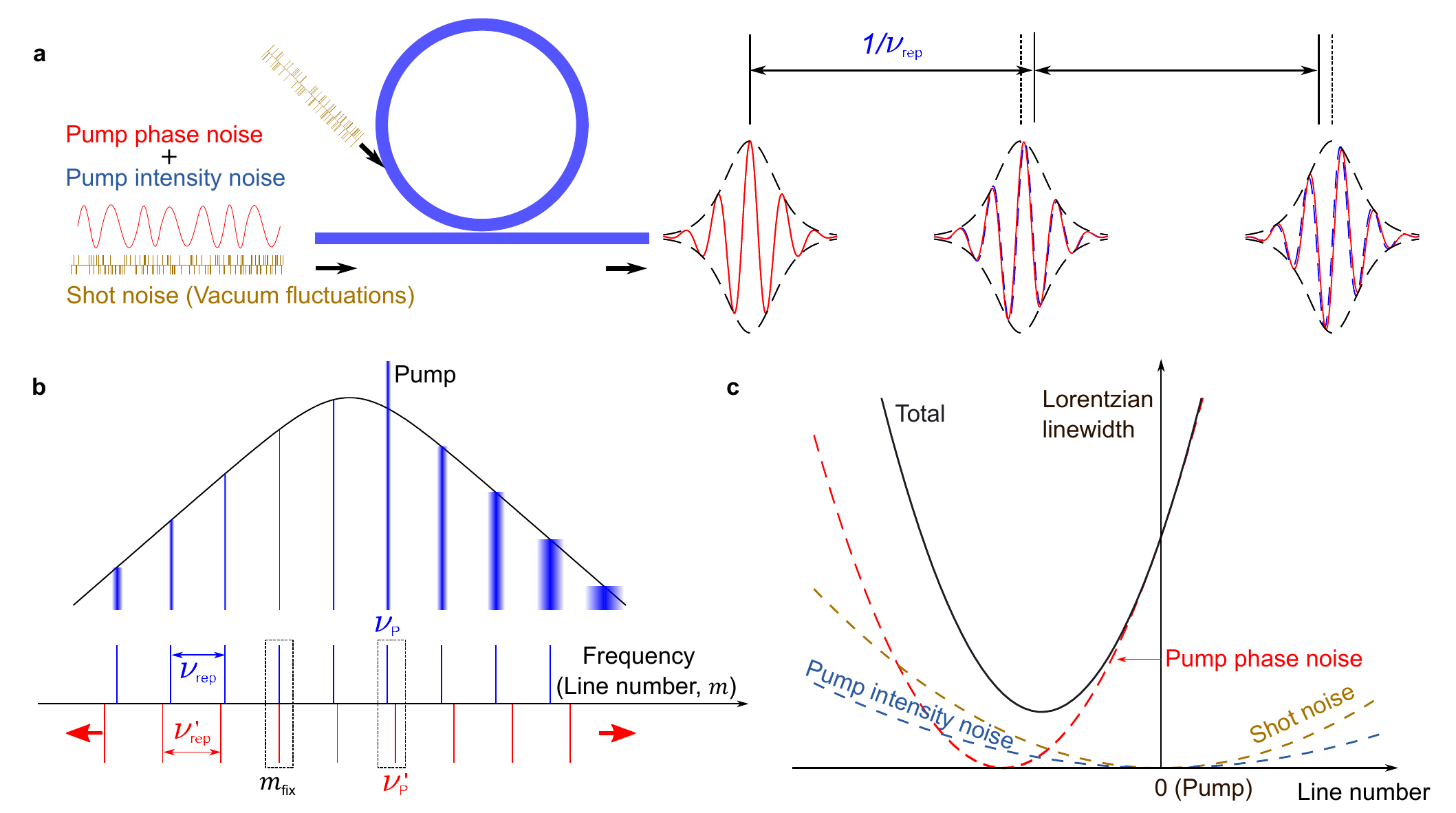}
\caption{ \textbf{Noise induced frequency fluctuations and optical linewidth narrowing in soliton microcombs}. \textbf{a} Soliton microcombs are generated by coupling a continuous-wave laser into a longitudinal mode of a high-Q microresonator. The phase and intensity noise of the pump and shot noise (vacuum fluctuations) cause timing jitter and pulse carrier-envelope offset fluctuations that result into a finite optical linewidth of the frequency comb lines.  \textbf{b} Because of the Raman self-frequency shift, variations in pump's laser  frequency to the blue side result in an increase of the soliton repetition rate, and vice versa to the red side. The coupling between pump frequency and repetition rate thus results in the existence of a so-called "fixed point" in the comb spectrum, that is a comb line that is most resilient to the fluctuations of the pump's frequency noise. This translates into a Lorentzian linewidth distribution with line number whose minimum can be located far away from the pump. \textbf{c} The shot noise and pump intensity noise affect directly the timing jitter of the soliton pulse train, which results into a linewidth distribution symmetrically located around the pump. These three effects together set the lowest achievable Lorentzian linewidth of soliton microcombs. }
\label{fig:1}
\end{figure*}

In this section, we analyse from a theoretical perspective the Lorentzian linewidth of soliton microcombs. Concretely, we analyze the flat frequency noise component of the PSD of the different comb lines. In general, both the pump laser’s phase noise and intensity noise have an impact on the Lorentzian linewidth of the comb lines. In addition, we shall consider shot noise affecting both the pump laser 
and the cavity, as an additive zero-point fluctuation field \cite{matsko2013timing} (see Fig. 1a). TRN leaves its footprint in the low-offset frequency region of the PSD \cite{huang2019thermorefractive}. As a result, it will influence the effective linewidth, an aspect that will be addressed in the latter sections.

We begin by considering the contribution of the frequency noise of the pump. The optical frequency of the $m$-th microcomb line $\nu_{\rm m}$ is determined by two degrees of freedom, i.e. the frequency of the pump laser $\nu_{\rm p}$ and  the repetition rate of the soliton microcomb $\nu_{\rm rep}$,
\begin{equation}
    {\nu_{\rm m}={\nu_{\rm p}}+m {\nu_{\rm rep}}},
    \label{eq1}
\end{equation} 
with the comb line number, $m$, counted from the pump. 
The underlying assumption in the elastic-tape model \cite{telle2002kerr} is that the noise sources will result in collective fluctuations of the comb lines. According to this, equation (\ref{eq1}) indicates that the linewidth of the pump would be faithfully imprinted on all other comb lines if the repetition rate were fixed. However, in soliton microcombs, due to the existence of intrinsic intrapulse Raman scattering \cite{karpov2016raman} and dispersive-wave recoil \cite{brasch2016photonic,matsko2016optical, yang2016spatial}, the pump phase  noise will also affect the repetition rate.  The repetition rate can be written as \cite{yi2017single}
\begin{equation}
 {\rm \nu_{\rm rep}}=\frac{1}{2\pi}[D_1+\frac{D_2}{D_1}(\Omega_{\rm Raman}+\Omega_{\rm Recoil})],
\label{eq2}
\end{equation} 
where $D_1/2\pi$ is the cavity's free spectral range (FSR) at the pump mode, and $D_2=- D^2_1 \beta_2/\beta_1 $ with $\beta_1$ and $\beta_2$ being the first- and second-order coefficients of the Taylor expansion of the mode propagation constant $\beta$ \cite{agrawal2000nonlinear}. $\Omega_{\rm Raman}$ and $\Omega_{\rm Recoil}$ denote the shift of the carrier frequency due to Raman scattering, i.e. the Raman self-frequency shift \cite{karpov2016raman} and dispersive-wave recoil \cite{yi2017single}, and both are
 functions of the detuning between the pump's cavity mode and the laser frequency, ($\rm \nu_c-\nu_p$)  \cite{yi2017single,bao2017soliton}. For simplicity, we only consider the Raman term in the main text, while the effect of dispersive-wave recoil is discussed in Supplementary Note 1.

According to  equations. (\ref{eq1}) and (\ref{eq2}), the frequency change of the $m$-th comb line $\delta \nu_{ m}$ induced by that of the pump $\delta \nu_{\rm  p}$ can be written as
\begin{equation}
    \delta \nu_{\rm m}=\delta{\nu_{\rm p}}(1+m \frac{d\nu_{\rm rep}}{d\nu_{\rm p}}).
    \label{eq3}
\end{equation} 
In the following, we will characterize the frequency noise of the comb lines by means of the frequency noise PSD, $S_{\rm \Delta\nu,m}(f)$, (sometimes called FM noise). 
When flicker noise and other technical sources are neglected, the pump frequency noise PSD $S^{\rm p}_{\Delta\nu}(f)$ is characterized by a constant level and has a simple relation with its fundamental phase noise PSD $S^{\rm p}_{\Delta\phi}$ as 
\begin{equation}
   S^{\rm p}_{\Delta\nu}(f)=f^2S^{\rm p}_{\Delta\phi}(f)= S_0.
    \label{eq4}
\end{equation}
The full-width at half maximum (FWHM) accounts for the Lorentzian linewidth, $\Delta\nu_{\rm p}=\pi S_0$ \cite{kikuchi2012characterization,liehl2019deterministic}. From equation (\ref{eq3}), the frequency noise PSD of the $m$-th comb line, $S_{\Delta\nu,{\rm m}} (f)$, can be directly linked to the frequency noise PSD of the pump as
\begin{equation}
   S_{\Delta\nu,{\rm m}}(f)=S^{\rm p}_{\Delta\nu}(f)
   (1-\frac{m}{m_{\rm fix}})^2.
   \label{eq5}
\end{equation}
where $m_{\rm fix}=-({d\nu_{\rm rep}}/{d\nu_{\rm p}})^{-1}$ corresponds to the pump phase-noise fixed point \cite{telle2002kerr}. Equation (\ref{eq5}) indicates that the Lorentzian linewidth of the microcomb induced by the pump linewidth follows a parabolic distribution with  line number, with a minimum value reaching zero at the fixed point, see Fig. \ref{fig:1}. For soliton microcombs affected by the Raman self-frequency shift, the fixed point appears to the red side of the pump because the repetition rate increases when the pump frequency increases. While the dispersive-wave recoil induced by  third-order dispersion or mode coupling could modify the location of the fixed point (Supplementary Note 1). Importantly, the elastic-tape model applies in these two cases, in spite of different physical mechanisms coupling the repetition rate with detuning.

In addition to the pump phase noise, the shot noise  could also introduce frequency noise into the soliton microcomb. 
In the studies of supercontinuum generation, shot noise poses fundamental limitations in the achievable spectral coherence  \cite{corwin2003fundamental,dudley2006supercontinuum}. In soliton microcombs, the shot noise hardly affects the pump frequency, however it sets a fundamental-limited timing jitter \cite{matsko2013timing,bao2021quantum, jeong2020ultralow,jia2020photonic}. As the repetition rate is mainly determined by $D_1$, the timing jitter PSD, $S^{\rm Q}_{\rm tm}(f)$, affects the optical frequency noise of the comb lines $S_{\rm \Delta\nu,m}(f)$ through
\begin{equation}
    S_{\rm \Delta\nu,m}(f)=m^2f^2{D_1}^2S^{\rm Q}_{\rm tm}(f).
    \label{eq6}
\end{equation}
With the fact $S^{\rm Q}_{\rm tm}(f) \propto 1/f^2$ if $f\ll{2\pi \nu_{\rm p}}/{Q}$ \cite{matsko2013timing,bao2021quantum}, one can see that the shot noise leads to a line-number-dependent white optical frequency noise. Since the effect arising from  spontaneous Raman scattering is usually much weaker than the shot noise \cite{corwin2003fundamental}, it is ignored here.  

In addition to the above two fundamental noise sources with quantum origin, in practice the intensity noise of the pump could also introduce frequency noise into the soliton microcomb because the repetition rate can be modified by the pump power via the Raman and dispersive-wave emission, but with little influence on  the pump frequency. The frequency noise PSD induced by the relative intensity noise (RIN) of the pump  $S^{\rm p}_{\rm RIN}(f)$ 
 can be  written as 
\begin{equation}
    S_{\Delta\nu,\rm m}(f)=m^2f^2{D_1}^2S^{\rm RIN}_{\rm tm}(f).
    \label{eq7}
\end{equation}
where the RIN-noise-induced timing jitter $S^{\rm RIN}_{\rm tm}(f)\propto S^{\rm p}_{\rm RIN}(f)/f^2$ if $f\ll{2\pi \nu_{\rm p}}/{Q}$. Note that the frequency dependence of the $S_{\rm RIN}(f)$ may be dominated by flicker noise. However, only its white frequency component is accounted for in the Lorentzian linewidth.
As illustrated in Fig. \ref{fig:1}{c}, the intensity noise introduces a  parabolic distribution of the Lorentzian linewidth whose center is located at the pump mode.

As the above three  noise sources (pump's intensity and phase noise, and shot noise) are independent from each other, their individual contribution to the linewidth can be added together. Thus, the Lorentzian  linewidth of the $m$-th line of a single-soliton microcomb can be expressed as
\begin{eqnarray}
    \Delta\nu_{\rm m}&=& \Delta\nu_{\rm p}(1-\frac{m}{m_{\rm fix}})^2
  + m^2(\Delta\nu_{\rm RIN}+\Delta\nu_{Q})
    \label{eq8}
\end{eqnarray}
with $\Delta\nu_{\rm RIN}=\pi {D_1}^2f^2S^{\rm RIN}_{\rm tm}(f)$ and  $\Delta\nu_{Q}=\pi {D_1}^2f^2S^{\rm Q}_{\rm tm}(f)$. 
Equation (\ref{eq8}) illustrates that because of the timing jitter induced by shot noise and intensity noise, the comb line with the minimum Lorentzian linewidth (we term it here the \textit{quiet mode}) is no longer the  phase-noise fixed point, and its location appears closer to the pump but remains always to the  longer-wavelength side (see Fig. \ref{fig:1}c). It is instructive to manipulate further equation (\ref{eq8}) and calculate the reduction of the linewidth at the quiet mode, $\Delta\nu_{\rm min}$, relative to the pump's Lorentzian linewidth
\begin{eqnarray}
   \frac{\Delta\nu_{\rm p}}{\Delta\nu_{\rm min}} &=& 
  {1+\frac{1}{m_{\rm fix}^2} \frac{\Delta\nu_{\rm p}}{\Delta\nu_{Q}+\Delta\nu_{\rm RIN}}}.
    \label{eq9}
\end{eqnarray}
Equation (\ref{eq9}) indicates that the relative reduction in linewidth is more prominent for pump lasers with larger Lorentzian linewidths. This observation allows for decreasing the Lorentzian linewidth of a coherent oscillator by performing frequency translation to the quiet mode with the aid of a soliton microcomb, an aspect that is addressed experimentally in the Supplementary Note 2.\\

\begin{figure*}[th]
\centering
\includegraphics[width=\linewidth]{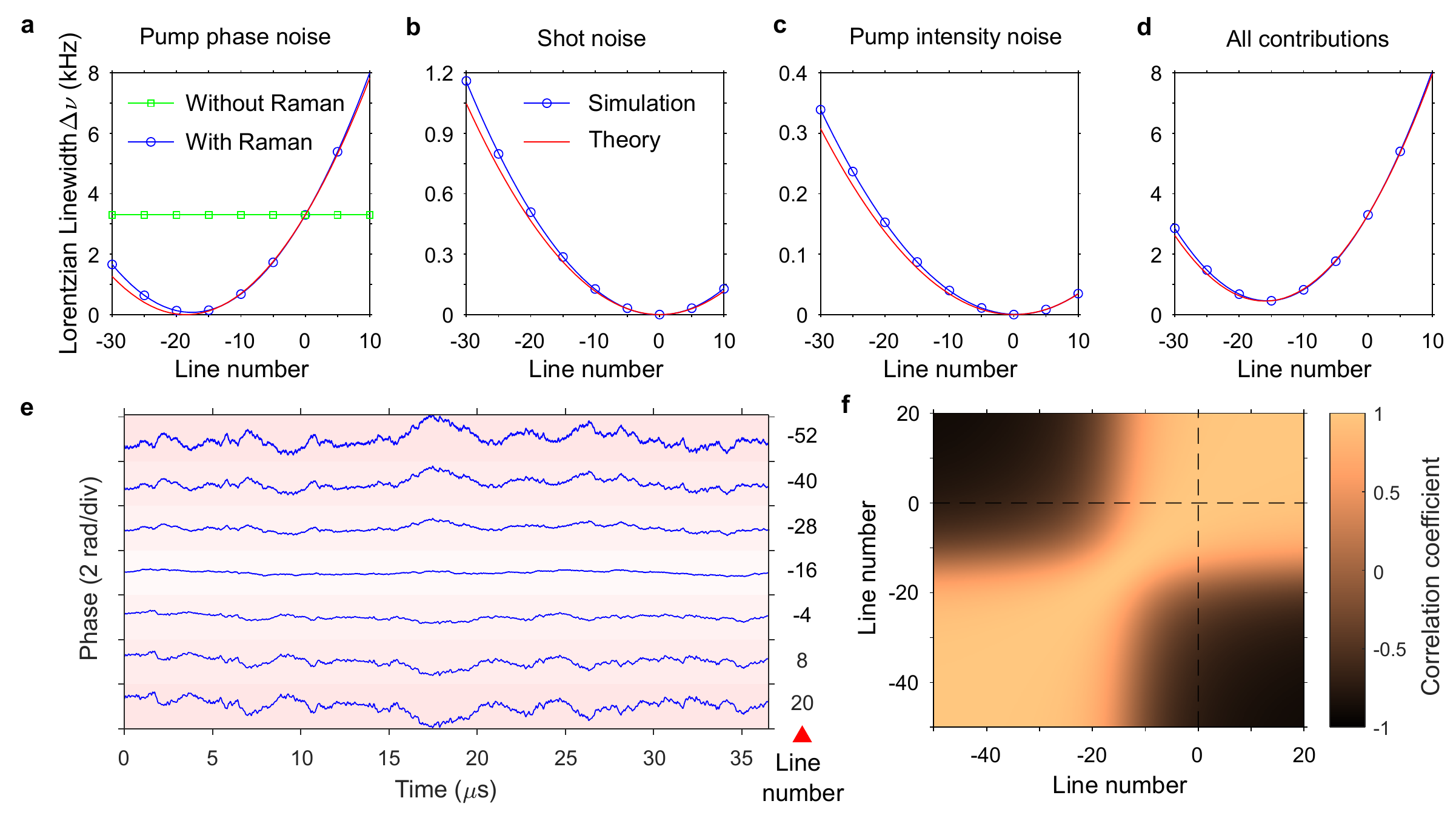}
\caption{\textbf{Numerical study of the Lorentzian linewidth of soliton mirocombs.} The Lorentzian linewidth of comb lines induced by pump phase noise, shot noise, pump intensity noise and when all noise sources are included, from \textbf{a} to \textbf{d}. The squares and circles denote the simulated data.  The theoretical results in \textbf{a}-\textbf{d} (red lines) are calculated according to equations (\ref{eq5}) - (\ref{eq8}).  \textbf{e},  Phase fluctuations of selected comb lines. \textbf{f}, Phase correlation among comb lines, computed according to equation (\ref{eq10}).}
\label{fig:2}
\end{figure*}

\noindent\textbf{Optical phase noise dynamics and numerical simulations.}   To test the validity of the previous theoretical analysis and gain a better understanding, we conduct a series of numerical simulations based on the Ikeda map \cite{ikeda1979multiple,hansson2016dynamics}. The parameters of the simulations are chosen to match the  characteristics of our silicon nitride (SiN) microresonators and are detailed in the Methods section.
The simulations also allow us studying the influence of the individual contribution of the different noise sources on the dynamics of the Lorentzian linewidth of soliton microcombs. We begin the analysis by considering the optical phase noise of the pump only (Fig. \ref{fig:2}a). In absence of Raman self-frequency shift, for a coherent dissipative soliton state, the phase noise is transduced equally among the comb lines, indicating a tight phase locking with the pump. However, when the Raman term is included, fluctuations in pump frequency are transduced into repetition rate changes. The repetition rate increases with detuning as a result of the Raman self-frequency shift, resulting in a fixed point to the red side of the pump characterized by near zero frequency fluctuations in theory and validated by the simulations. 

As anticipated in the previous section, both shot noise and intensity noise prevent from attaining a near zero Lorentzian linewidth at the fixed point. This is analysed in the numerical simulations presented in Fig. \ref{fig:2}b and c. Since shot noise and intensity noise do not modify the frequency of the pump but the repetition rate of the soliton microcomb, the distribution of the linewidth is symmetric with respect to the pump and follows a parabolic profile, in agreement with the theoretical analysis presented in the previous section, cf. equations. (\ref{eq6}) and (\ref{eq7}). Note that the Lorentzian linewidth induced by shot-noise timing jitter in our SiN microcomb is more than one order of magnitude higher than what has been previously reported for silica microcombs \cite{bao2021quantum}, mainly due to the larger nonlinear coefficient of the SiN cavity mode.  Hence, the shot noise sets a relatively high bound to the lowest achievable Lorentzian linewidth in this platform.

Figure  \ref{fig:2}d shows the simulation results when all noise contributions are added together. The Lorentzian linewidth based on the theoretical model (equation (\ref{eq8})) is plotted for comparison. The agreement between the theory and simulation indicates the elastic tape models can account for the most salient features of the optical phase noise dynamics in soliton microcombs. The slight deviation may arise from the fact that the elastic tape model implicitly assumes an instantaneous response in the two degrees of freedom of the comb, whereas in reality, there is an intrinsic latency in the system that can cause the comb to fluctuate in more than two degrees of freedom.

\begin{figure*}[ht]
\centering
\includegraphics[width=\linewidth]{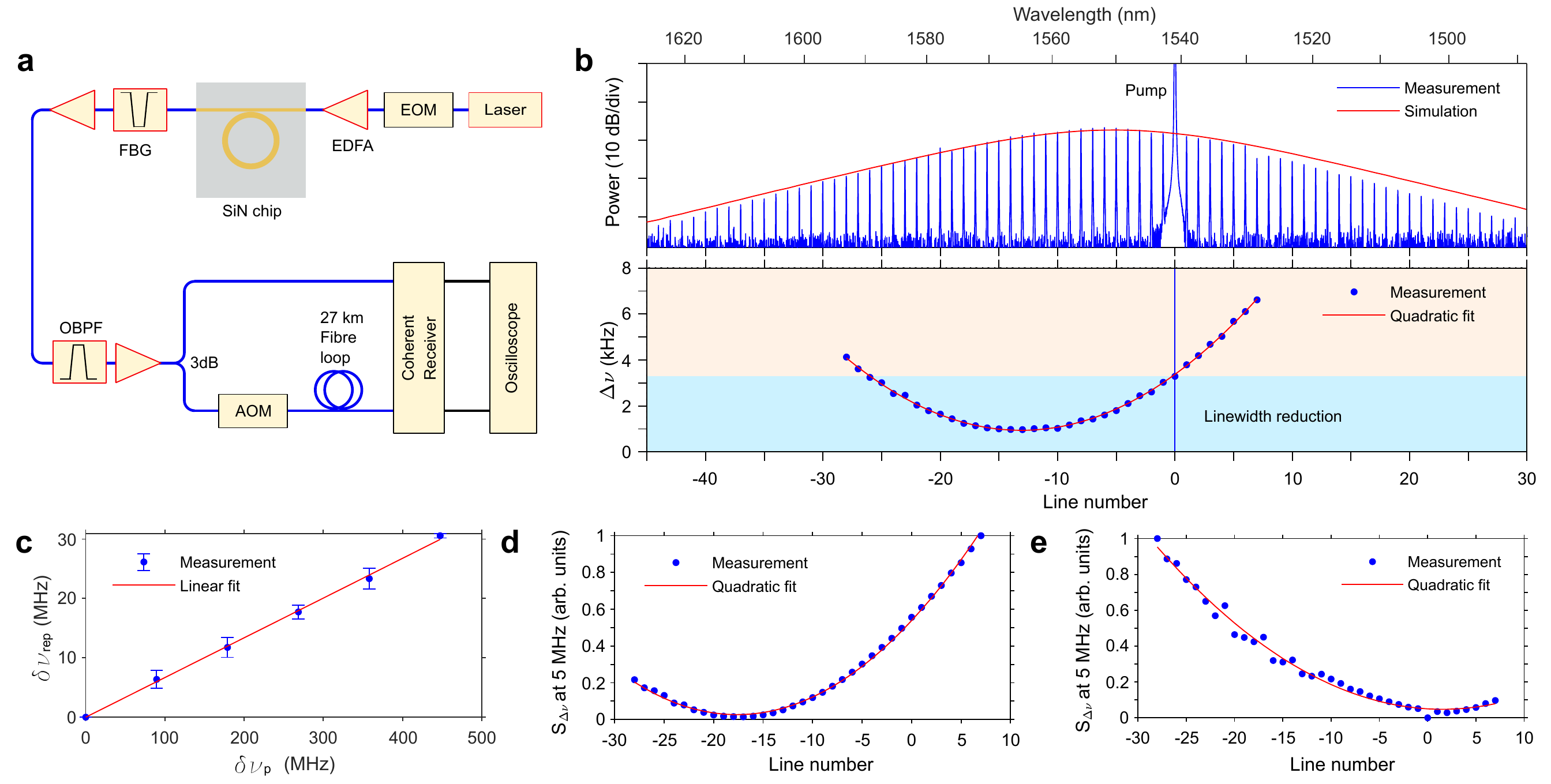}
\caption{\textbf{Experimental study of the Lorentzian linewidth of silicon nitride soliton microcombs.} \textbf{a} Setup for linewidth measurement. EOM, (Phase or Intensity) Electro-optical modulator; EDFA, Erbium-doped fibre amplifier; FBG, Fibre Bragg grating for pump rejection; OBPF, Optical band-pass filter; AOM,  Acousto-optic modulator.  \textbf{b} Optical spectrum of the single-soliton microcomb and the measured Lorentzian linewidth of comb lines according to the mean value of $S_{\rm \Delta\nu,m}(f)$ at high offset frequencies ranging 3 to 5 MHz.  \textbf{c}  Measured dependence of soliton microcomb repetition rate change on the relative pump frequency. The error bar stands for the standard deviation of three measurements. \textbf{d} and \textbf{e} The height of frequency noise PSD at 5 MHz, originating from  applying a phase or intensity modulation signal to the pump, respectively.}
\label{fig:3}
\end{figure*}

We end this section by analyzing the phase noise dynamics of the individual line components when all noise sources are included. Specifically, we compute the phase noise of  individual frequency lines, see  Fig. 2e. Clearly, the lines close to the quiet mode display reduced phase noise, with a standard deviation smaller than the inherent phase noise of the pump laser. Remarkably, the quiet mode stands out as a mirror symmetry point in the comb, whereby lines symmetrically located around it attain identical Lorentzian linewidth but anti-correlated phase noise. A similar behaviour has been observed before for electro-optic frequency combs, with the key difference that the fixed point corresponds there to the pump laser frequency \cite{lundberg2020phase,brajato2020bayesian}. This observation can be further quantified with the aid of the Pearson's correlation coefficient:
\begin{equation}
    C(m,n)=\frac{ cov(\phi_{\rm m},\phi_{\rm n})}{\sigma_{ \phi_{\rm m}}  \sigma_{ \phi_{\rm n}}},
    \label{eq10}
\end{equation}
where $\phi_{\rm m}$ denotes the sampled phase  for the $m$-th comb line. ${cov}$  is the covariance and $\sigma _{\rm X}$ is the standard deviation of $X$. The result is  plotted in Fig. \ref{fig:2}{f}. One can see that the phases of comb lines at the same side of  $m=-16$ are highly correlated, and those on opposite sides  anti-correlated, which can be explained with the elastic-tape model \cite{liehl2019deterministic}.\\

\noindent\textbf{Experiments with SiN microcombs.} We present  experimental results of the distribution of the frequency noise and corresponding Lorentzian linewidth in a soliton microcomb implemented in a silicon nitride microresonator pumped by a narrow-linewidth external cavity tunable diode laser (Santec TLS 710). The setup is shown in Fig. \ref{fig:3}{a}. The FSR and the average intrinsic Q factor of the SiN microresonator are $227.5 $ GHz and $1.16\times 10^7$. The pump laser is amplified by an EDFA. The power coupled into the bus waveguide is 120 mW. Other parameters and experimental details are described in the Methods.
The soliton microcomb generation is enabled through fast thermo-optic tuning via an integrated metallic heater \cite{joshi2016thermally}. 
The optical spectrum of the generated single-soliton microcomb is shown in Fig. \ref{fig:3}{b}. After  attenuation of the pump line with a notch filter, the comb is amplified with either a C- or L-band EDFA. Then each comb line is filtered out separately and amplified to $\sim$ 20 mW. Subsequently, the phase and frequency noise of each comb line is measured through a self-heterodyne measurement technique implemented with the aid of a coherent receiver (Neophotonics, 100 Gbps micro-ICR) \cite{kikuchi2012characterization}.

The measured Lorentzian linewidth for the comb lines from $m= -28$ to $m=7$ is shown in Fig. \ref{fig:3}{b}, which is obtained based on the average value at high offset frequencies (3-5 MHz) of the frequency noise PSD (Supplementary Note 3) \cite{jin2021hertz}. This region is chosen to avoid the contribution of flicker noise at low frequencies and the divergence at high frequencies caused by the white phase-noise component arising from both the optical amplifiers ASE noise and the thermal noise of the measurement system \cite{newbury2007low}. The results indicate that a significant portion of the comb lines around -13  (equivalent 1565 nm wavelength) display a Lorentzian linewidth that is in fact smaller than that of the pump. Note that the location of the quiet mode is close to the simulation in Fig. 2d. The line located at -13 achieves the smallest phase noise, with a Lorentzian linewidth of $\sim$1 kHz, corresponding to  more than a threefold reduction of the pump's value. The measurement is higher than the predictions ($\sim$0.5 kHz) provided by Fig. 2d. We believe the  discrepancy is partly due to the TRN in the microresonator, which has a non-negligible contribution in the frequency range used for computing the Lorentzian linewidth \cite{jin2021hertz}. Similar to the effect of the intensity noise, TRN leads to timing jitter or repetition rate change of soliton microcombs. Other effects that might contribute to the frequency noise dynamics include a frequency-dependent Q factor \cite{matsko2015noise} or avoided mode crossings.

We continue with a quantitative analysis of the location of the fixed point by measuring the change of the soliton's repetition rate with  pump frequency, cf equation (5). Specifically, the repetition rate of the soliton is measured by electro-optic downconversion \cite{del2012hybrid} as the pump is set at different values, which are measured with a wavelength meter. The results of this measurement are presented in Fig. \ref{fig:3}{c}. The slope of the variation is positive, explaining why the fixed point appears to the red side of the pump, with the estimated fixed point $m_{\rm fix}=-({d\nu_{\rm rep}/d\nu_{\rm p}})^{-1}=-15$. However, this measurement provides limited insight into the phase-to-phase transduction from the pump, because the temperature of the resonator is not the same when the detuning is changed. Such measurements are further addressed in the following experiments. We modulated the pump laser with either a phase or intensity electro-optic modulator (EOM) at an arbitrary specific single radio frequency before getting amplified by an EDFA. This  introduces spikes in the $S_{\rm \Delta\nu,m}(f)$ at the corresponding  modulation frequency, at levels about 3 orders higher than the  noise background. The measurement results are presented in Figs. \ref{fig:3}{d} and \ref{fig:3}{e}, corresponding to the phase and intensity modulation, respectively. The magnitude of the spike provides an indirect estimation of the influence of the pump's phase/amplitude noise on the Lorenztian linewidth of the corresponding comb line. One can see that while the intensity-to-phase noise transduction is symmetrically located around the pump,  the phase-to-phase distribution  attains its minimum at   $m_{\rm fix}=-18$, which is more in line with the simulations in Fig. \ref{fig:2}a.  Moreover, it is interesting to note that compared to the phase modulation spike, the intensity modulation spike fluctuates more. This can be explained as intensity-to-phase noise transduction is more sensitive to the detuning which may change slightly during the measurement. \\

\begin{figure*}[ht]
\centering
\includegraphics[width=\linewidth]{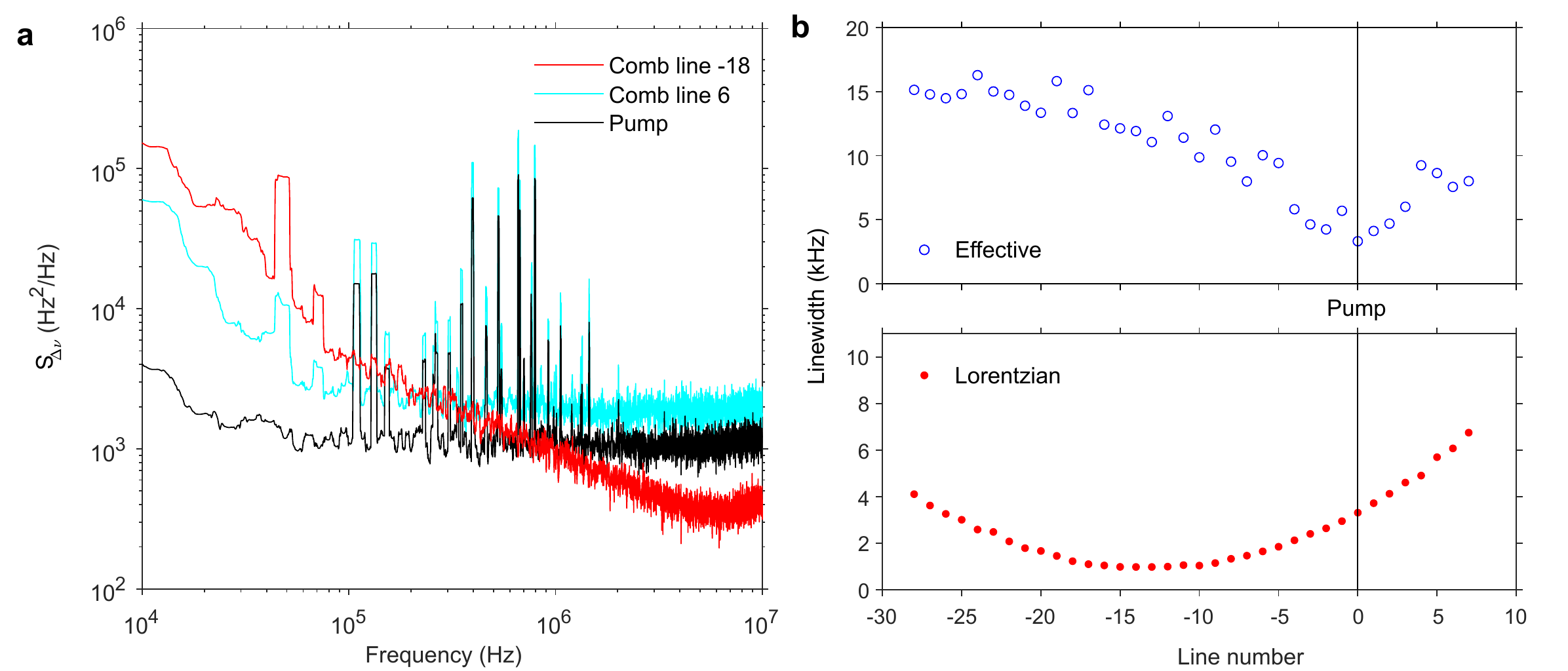}
\caption{\textbf{Experimental comparison between effective and Lorentzian linewidth.} \textbf{a} The frequency noise power spectral density $S^{\rm p}_{\Delta\nu}(f)$ for two different comb lines. At low offset frequencies, all comb lines display higher frequency noise than the pump due to the TRN, while at high offset frequency range, some comb lines close to the quiet mode may have lower frequency noise than the pump. \textbf{b} Comparison between  Lorentzian and effective linewidth. Because of the low-frequency noise, the effective linewidth is higher than the Lorentzian linewidth, and the linewidth distribution appears symmetrically located with respect to the pump. }
\label{fig:4}
\end{figure*}

\noindent\textbf{Lorentzian linewidth and effective linewidth.} 
The above analysis focused on the Lorentzian linewidth and ignored the low offset frequency noise contribution. In microresonators with small cavity volumes, the TRN has a significant contribution  on the frequency noise PSD, specially at low frequencies \cite{gorodetsky2004fundamental,matsko2007whispering,huang2019thermorefractive}.  This noise source causes fluctuations in the location of the longitudinal modes and  repetition rate of the soliton through the detuning parameter \cite{drake2020thermal}. As a result, the PSD of the comb lines get an additional contribution at low frequencies, see Fig. \ref{fig:4}a (and Supplementary Fig. 4). To account for such non-white frequency noise, it is useful to calculate the effective linewidth of the comb lines $\Delta{\nu}_{\rm m}^{\rm eff}$ using the following definition
\cite{hjelme1991semiconductor}:
\begin{eqnarray}
   \int_{\Delta{\nu}_{\rm m}^{\rm eff}}^{\infty} \frac{S_{\Delta\nu,{\rm m}}(f)}{f^2} df=\frac{1}{\pi}.
 \label{eq11}
\end{eqnarray}

The results are displayed in Fig. \ref{fig:4}b (see also Supplementary Note 4), and compared to the results in Fig. \ref{fig:3}b, reproduced again for convenience. Clearly, the non-white frequency noise region (or flicker noise) has a dominant contribution in the value of the effective linewidth. The distribution is symmetrical with respect to the pump. This is expected, and in line with previous reports \cite{nishimoto2020investigation}.\\ 

\begin{figure}[h]
\centering
\includegraphics[width=\linewidth]{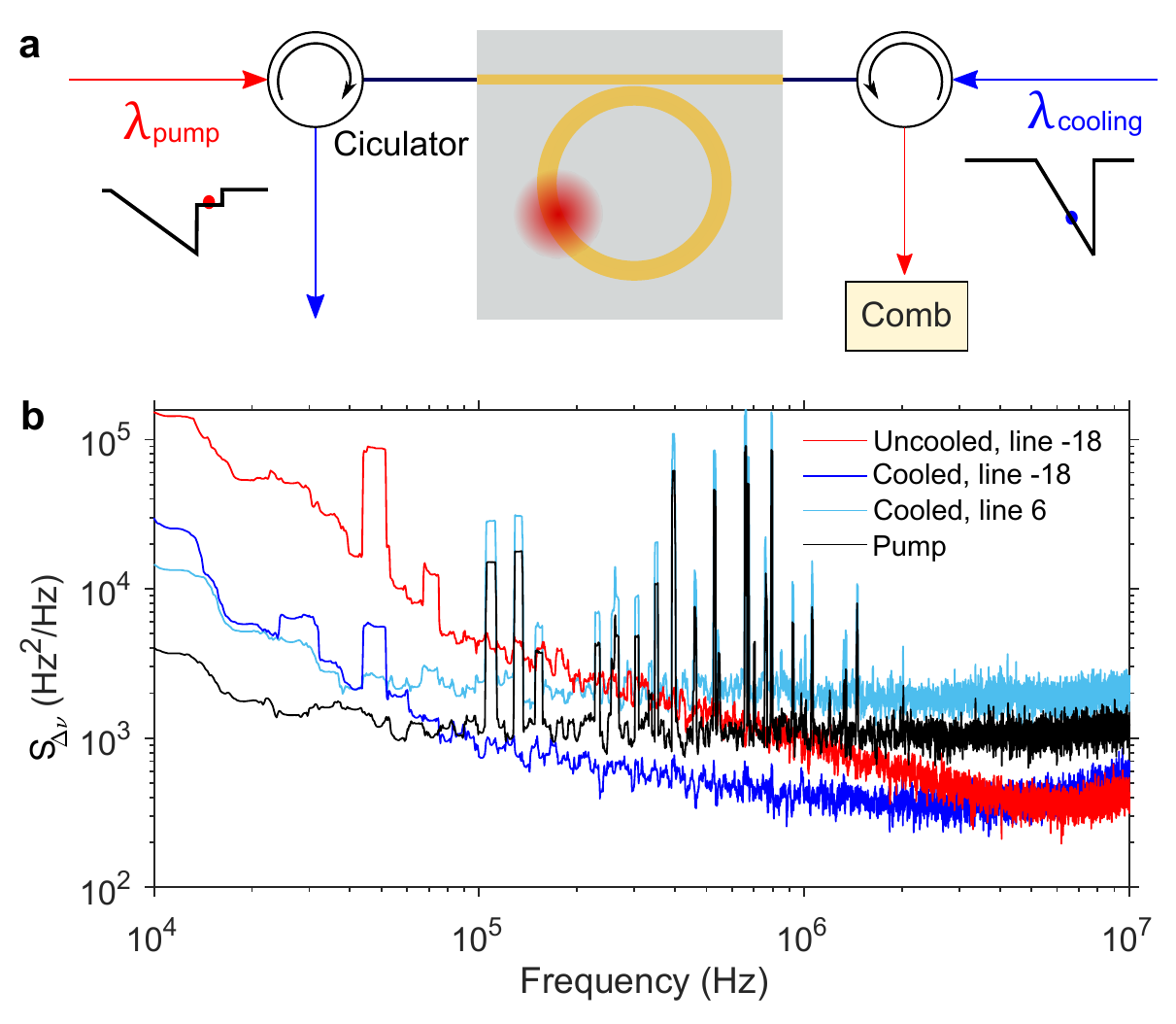}
\caption{\textbf{Frequency noise reduction with laser cooling.}  \textbf{a} The schematic of the setup. An auxiliary laser (blue arrow) opposite to the pump (red arrow) is injected into the microresonator and blue-detuned to a cavity mode with the same family as the pump mode. $S_{\Delta\nu,-18}(f)$ without (red) and with (blue) laser cooling are shown in  \textbf{b}. As reference,  the pump frequency noise PSD $S^{\rm p}_{\Delta\nu}(f)$ and $S_{\Delta\nu,6}(f) $ with laser cooling are shown as well.}
\label{fig:5}
\end{figure}

\noindent\textbf{Frequency noise reduction in soliton microcombs.} It is important to note that, the frequency noise reduction experienced by the comb lines nearby the fixed point takes place at all Fourier frequencies in the PSD. Not only the Lorentzian linewidth is reduced, but the low offset technical noise of the pump is also reduced according to equation (\ref{eq5}). This is clearly demonstrated in Fig. 4a, where the pump displays a set of spikes in the PSD within the range 100 kHz-1 MHz. The spikes are significantly damped for the comb line -18, which is very close to the fixed point (see Supplementary Figure 4c). This analysis points to the intriguing possibility of soliton microcombs to generate coherent oscillators on chip with a linewidth narrower than the pump itself – a feature that works more efficiently for broader linewidth lasers, cf equation (\ref{eq9}). An experimental demonstration is given in the Supplementary Figure 3. When TRN contributes, such a reduction is however masked at low offset frequencies, see Fig. 4a,  therefore, to fully capitalize on this characteristic, effective means to suppress TRN are needed, such as laser cooling \cite{drake2020thermal,sun2017squeezing,lei2022thermal,nishimoto2022thermal}, cavity dispersion engineering \cite{stone2020harnessing} or directly operating at cryogenic temperatures \cite{moille2019kerr}. 

In the  following, we investigate the laser cooling technique. 
The laser cooling is performed with an auxiliary laser coupled to the microresonator from the opposite direction, as schematically depicted in Fig. \ref{fig:5}{a}. The cooling laser (Toptica, CTL 1550)  has an on-chip power around 2 mW and is  tuned into a resonance close to 1550 nm and kept blue detuned. The measured  $S_{\rm \Delta\nu,m}(f)$ for comb line $m=-18$ is shown in Fig. \ref{fig:5}{b}. By applying the laser cooling technique, the frequency noise can be suppressed, but only up to a certain offset frequency. Take for example line $m=-18$, close to the fixed point. The laser cooling reduces up to an order of magnitude the frequency noise at low offset frequencies, but does not result in a further reduction in Lorentzian linewidth. This is more clearly observed when comparing the PSD of line $m=6$, which is far away from the fixed point and, in the presence of laser cooling, its white frequency noise plateau is still above the corresponding level of the pump. \

\noindent\textbf{Discussion} 

In summary, we have analysed the optical linewidth of soliton microcombs, including both the Lorentzian components and the low offset frequencies in the frequency noise spectral density. We discovered that the elastic-tape
model, found previously for mode-locked lasers, does apply
for soliton microcombs.  This is not a mere translation
of the model because in soliton microcombs the
pump laser is coherently added to the soliton spectrum, which results into unexpected findings. In particular, because of the Raman self-frequency shift and disperve-wave recoil, the model predicts the existence of a frequency region, potentially far away from the pump, where the soliton microcomb lines attain a Lorentzian linewidth below the one displayed by the pump laser. 

This inherent property of soliton microcombs for optical phase noise reduction could be used to generate coherent oscillators derived from broad linewidth lasers. In practice, however, it is the TRN of the cavity that contributes the most at low-offset frequencies in the frequency noise power spectral density, resulting in a dominant contribution in the effective linewidth of the comb lines. This effect can be notwithstanding damped with the aid of a cooling laser. Further reduction of the TRN would be more helpful for this purpose, which might be realized by making the cavity with athermal material or increasing its size. \\

\noindent\textbf{Methods}\\
\noindent\textbf{Resonator characteristics and operating condition.}
The SiN microring resonator is fabricated via subtractive processing as described in \cite{ye2019high}. The radius of the microresonator used for the experiment is 100 $\mu$m, and the corresponding FSR is about 227.5 GHz. The height and the width of the SiN waveguide are 750 nm and 2100 nm, respectively, which result in a group velocity dispersion coefficient of $\rm \beta_2=$-80 $\rm  ps^2/km$ for the $\rm TE_{00}$ mode.
The average intrinsic Q-factor of the sample is 11.6 million and the total Q-factor is  8 million. Lensed fibres are used for  coupling into and out of the on-chip SiN bus waveguide. The average coupling loss per facet is about 3.5 dB. \\

\noindent\textbf{Theory and simulation.}
To apply the theoretical results corresponding to  equations (\ref{eq5})-(\ref{eq7}) in Fig. \ref{fig:2}a-d,  the parameter $m_{\rm fix}$ was calculated using its definition, i.e., the derivative of repetition rate with respect to the pump's frequency. The repetition rate was attained by monitoring the speed of the soliton, whose temporal position $t_{\rm p}$ is calculated as \cite{paschotta2004noise}
\begin{eqnarray}
    t_{\rm p}=\frac{\int t |A(t)|^2dt}{\int|A(t)|^2dt}.
\end{eqnarray}
The shot-noise-induced and RIN-induced timing jitter PSDs $S^{\rm Q}_{\rm tm}(f)$ and $S^{\rm RIN}_{\rm tm}(f)$ were obtained from simulations since there is no analytical result  that includes the Raman self-frequency shift. 

The simulation is performed with the Ikeda map \cite{ikeda1979multiple}. A full roundtrip evolution is divided into two steps including (1) the coupling between bus waveguide and resonator, (2) nonlinear propagation in the resonator over its circumference.

The coupling between bus waveguide and resonator can be described as \cite{hansson2016dynamics}
\begin{eqnarray}
A_{\rm m+1}(0,\tau)=\sqrt{\theta}A_{\rm m}^{\rm in}+\sqrt{1-\theta}e^{i\phi_0}A_{\rm m}(L,\tau),
\end{eqnarray}
where $A_{\rm m}$ stands for the amplitude (normalized to power) of intracavity field of $m$-th round trip, $\phi_0 ={2\pi(\nu_{\rm p}- \nu_{\rm c})}/{\rm FSR}$, and $\theta = {2\pi \nu_{\rm p}}/({{\rm FSR}\times Q_{\rm ex}}) $ with $Q_{\rm ex}$ as the extrinsic quality factor. In each step, the pump phase noise can be included into $A_{\rm m}^{\rm in}$ by setting $A_{\rm m}^{\rm in}=\sqrt{P_{\rm in}}e^{i\phi_{\rm m}}$. The white pump frequency noise can be simulated by generating a time-dependent pump phase through
\begin{equation}
    \phi_{\rm m+1}=\phi_{\rm m}+\sqrt{2\pi\Delta{\nu_{\rm p}}/{\rm FSR}}\times \eta,
\end{equation}
where $\eta$ stands for a normally distributed random number.

The shot noise is treated via a semi-classical method described in \cite{paschotta2004noise,drummond2001quantum}. 
Specifically, we add a noise field   coupled to the resonator  and set its random amplitude $\delta A^{\rm in}(t)$ with statistics \cite{paschotta2004noise}
\begin{equation}
    <(\delta {A^{\rm in}}(t))^*\delta A^{\rm in}(t+\tau)>=\frac{h\nu_{\rm p}}{2}\delta(\tau).
\end{equation}
Here $h$ is the Planck constant. The coupling coefficient is set to $\sqrt{2\pi\nu_{\rm p}/(Q_{\rm ex}+Q_{\rm in})/{\rm FSR}}$, according to the fluctuation-dissipation theorem. This can be implemented in the split-step Fourier method with assignment of $\delta A^{\rm in} = ({\eta_1+i\eta_2})\sqrt{{h\nu_{\rm p}}\times N{\rm \times FSR}}/2$ in the fast time domain, where $\eta_{1(2)}$ stands for a normally distributed random number and $N$ is the number of discretization points.

To account for the frequency noise at the frequencies between 3-5 MHz contributed by the pump intensity noise, here we use the measured  relative intensity  noise (RIN) (${S}^{\rm p}_{\rm RIN}= -127.5 $ dBc/Hz) in this region for the simulation.    
To incorporate the intensity noise into the simulation, a fluctuation term $\delta P_{\rm in}$ is added into the input power $P_{\rm in}$ for each roundtrip satisfying
\begin{equation}
    \delta P_{m}= P_{\rm in} \sqrt{ S^{\rm p}_{\rm RIN} {\rm FSR/2}}\times \eta.
\end{equation}

At each  propagation step, the generalized nonlinear Schrödinger equation with (or without) inclusion of Raman scattering is solved \cite{agrawal2000nonlinear,dudley2010supercontinuum}:
\begin{eqnarray}
   \frac{\partial A}{\partial z}+  & &\frac{\alpha}{2} A-i\sum_{n=1}^3\frac{i^n\beta_{\rm n}}{n!}\frac{\partial^n A}{\partial t^n}  = i\gamma A(z,t)
   \nonumber \\
   & & \times\int_0^\infty R(t')|A(z,t-t')|^2dt'. 
\end{eqnarray}
Here $R(t)=(1-f_{\rm R})\delta(t)+f_{\rm R} h_{\rm R}(t)$, and $h_{\rm R}(t)=(\tau_1^{-2}+\tau_2^{-2})\tau_1 {\rm exp}({-t}/{\tau_2}){\rm sin}({t}/{\tau_1})$, and the parameters  have the same meaning as  \cite{agrawal2000nonlinear}. In the simulation, the parameters $\tau_1=15 $ fs, $\tau_2=120 $  fs, $f_{\rm R}=0.027$ are  used as they match the measured optical spectrum. The remaining  parameters  are directly measured or simulated and have the following values: $
P_{\rm in}=0.1$ W,
$\rm \gamma=0.9W^{-1}m^{-1}$, $Q_{\rm ex}=3.15\times10^7$, $Q_{\rm in}=1.35\times10^7$, $\Delta{\nu_{\rm p}}=3.3$ kHz, $\nu_{\rm c}-\nu_{\rm p}= 626$ MHz.

To extract the phase noise information from the optical modes, around 8 million roundtrips were simulated in order to have reliable statistics at low offset frequencies. The phases of selected modes (9 modes equally spaced from comb line  $m=-30$ to $m=10$) were recorded  every 512 roundtrips. With the recorded phases, we computed  $ S_{\rm \Delta\nu,m}(f)$ and the Lorentzian linewidth for each comb line  according to their definitions.\\

\noindent\textbf{Frequency noise measurement and effective linewidth.} With a coherent receiver, the complex amplitude of the beat between the laser and its delayed self is measured. Using the setup shown in Fig. \ref{fig:3}\textbf{a}, the phase difference at two times of the laser under test can be directly extracted:
\begin{equation}
   \Delta\phi (t)= \phi(t)-\phi (t-T),
     \label{eqc1}
\end{equation}
where $T$ stands for the time delay caused by a fiber loop with length of 27 km, as shown in Fig. \ref{fig:3}a. An acousto-optic modulator driven at 27 MHz is used to shift the measured signal out of baseband. 

According to Fourier analysis, we have 
\begin{equation}
|\mathscr{F}{\Delta\phi(t)}|^2=|\mathscr{F}{\phi(t)}|^2\times 4{\rm sin}^2(\frac{2\pi fT}{2}).
     \label{eqc2}
\end{equation}
Therefore, the $S_{\Delta \nu}(f)$ can be calculated by:
\begin{equation}
S_{\Delta\nu}(f)=\frac{f^2|\mathscr{F}{\Delta\phi(t)}|^2}{4{\rm sin}^2({\pi fT})}.
\label{eqc3}
\end{equation}
In practice, to avoid the divergence point in equation (\ref{eqc3}), one could do an average for equation (\ref{eqc2}). Noting  ${\rm <sin^2(x)>}=1/2$, we reach an approximated expression for $S_{\Delta\nu}(f)$:

\begin{equation}
S_{\Delta\nu}(f)=\frac{1}{2}<{f^2|\mathscr{F}{\Delta\phi(t)}|^2}>_{\rm T},
\label{eqc4}
\end{equation}
where $<.>$ stands for  the average over a period  $ T$. 

The effective linewidth was obtained according to equation (\ref{eq10}), by fitting the measured $\rm S_{\Delta\nu,m}(f)$  with a function $a+b/(f+c f^2)$. To emphasize the generality of the definition of the effective linewidth, the spikes (as shown in Fig. \ref{fig:5}) in the power spectral density were filtered out. \\

\noindent\textbf{Data availability} The data used in this work can be accessed from zenodo https://doi.org/10.5281/zenodo.6523268 \\

\noindent\textbf{Code availability} The simulation code may be available on reasonable request.

\bibliography{ref}

\noindent\textbf{Acknowledgments}\\
The devices demonstrated in this work were fabricated at Myfab Chalmers. We acknowledge funding support from the European Research Council (ERC, CoG GA 771410) and the Swedish Research Council (2015-00535, 2016-06077, 2020-00453).\\

\noindent\textbf{Author contributions}
F.L conducted the experiment with input from Z.Y.,Ó.B. H; F.L. provided the theoretical model and numerical simulations with input from Ó.B. H; Z.Y. and M.G. designed and fabricated the SiN devices; A.F. developed the phase noise measurement system.  All authors discussed the results. F.L and V. T.-C.  prepared the manuscript with input from all co-authors. V.T.-C. supervised this project.

\noindent\textbf{Additional information}\\
\textbf{Competing interests:} The authors declare no competing financial interests.

\section*{\textbf{Supplementary Note 1: Influence of the dispersive-wave recoil on the location of the fixed point}}

In the main text, we have discussed the transduction of the pump frequency noise into repetition rate noise induced by the Raman self-frequency shift.  In this section, we study the influence of dispersive-wave recoil. The dispersive wave could be introduced by third-order or higher-order dispersion of the microresonator or the mode-coupling induced mode shift [1-3]. First, we analyse the influence of the third-order  dispersion (TOD) numerically. As shown in Supplementary Fig. 1a, the location of the fixed point can be changed slightly with the TOD, with all the other parameters being equal. More interestingly, in absence of Raman, TOD can lead to linewidth narrowing as well, and the fixed point could be located either to the blue or red side depending on the sign of  $\beta_3$. However, the TOD is typically  so weak that the fixed point appears outside of the comb spectrum, as shown in Supplementary Fig. 1b. 


\begin{figure*}[h]
\centering
\includegraphics[width=\linewidth]{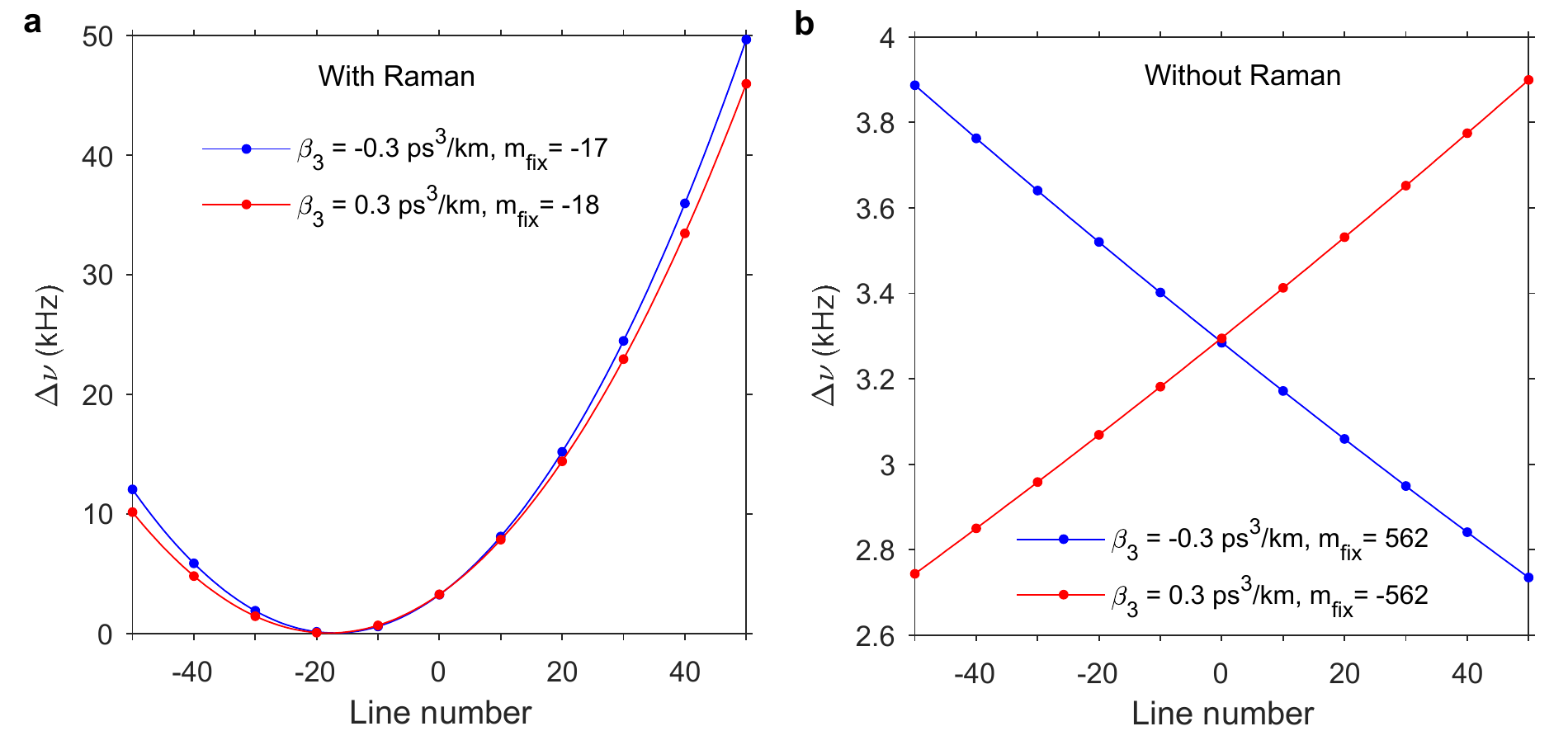}
\renewcommand{\figurename}{{Supplementary Figure}}
\caption{The linewith of comb lines induced by pump phase noise with existence of third-order dispersion (TOD). The TOD can modify the linewidth of comb lines, when the Raman is included \textbf{a} or excluded \textbf{b}.}
\label{fig:s1}
\end{figure*}

It has been demonstrated that the mode interaction induced dispersive-wave recoil could balance the Raman self-frequency shift, leading to quiet point, where the repetition rate is insensitive to variation of the pump-resonance detuning [4].  However, the mode interaction is often uncontrollable in single ring cavities. Here, we consider a microcomb generated in a photonic molecule configuration [5], see Supplementary Fig. 2a. The width and height of the two microrings are 1600 nm and 600 nm respectively, resulting in a group velocity dispersion coefficient of $\rm \beta_2=$100 $\rm  ps^2/km$ for mode $\rm TE_{00}$. The FSRs for the two rings are 100.79 GHz and 99.83 GHz respectively. The two rings feature similar intrinsic Q (5 million) and extrinsic Q (6.5 million) factors.  The gap between the two rings is 500 nm, which results in mode coupling between two rings. The coupling induced mode splitting is 2$\pi\times$607 MHz. To generate the soliton, we pump a supermode at 1569.44 nm with on-chip power around 40 mW, with similar operation to [5]. In addition to the pump mode, many other pairs of longitudinal modes from the two microrings are also coupled, leading to a complicated dispersion profile and frequency spectrum, see Supplementary Fig. 2b. 


\begin{figure*}[hb]
\centering
\includegraphics[width=\linewidth]{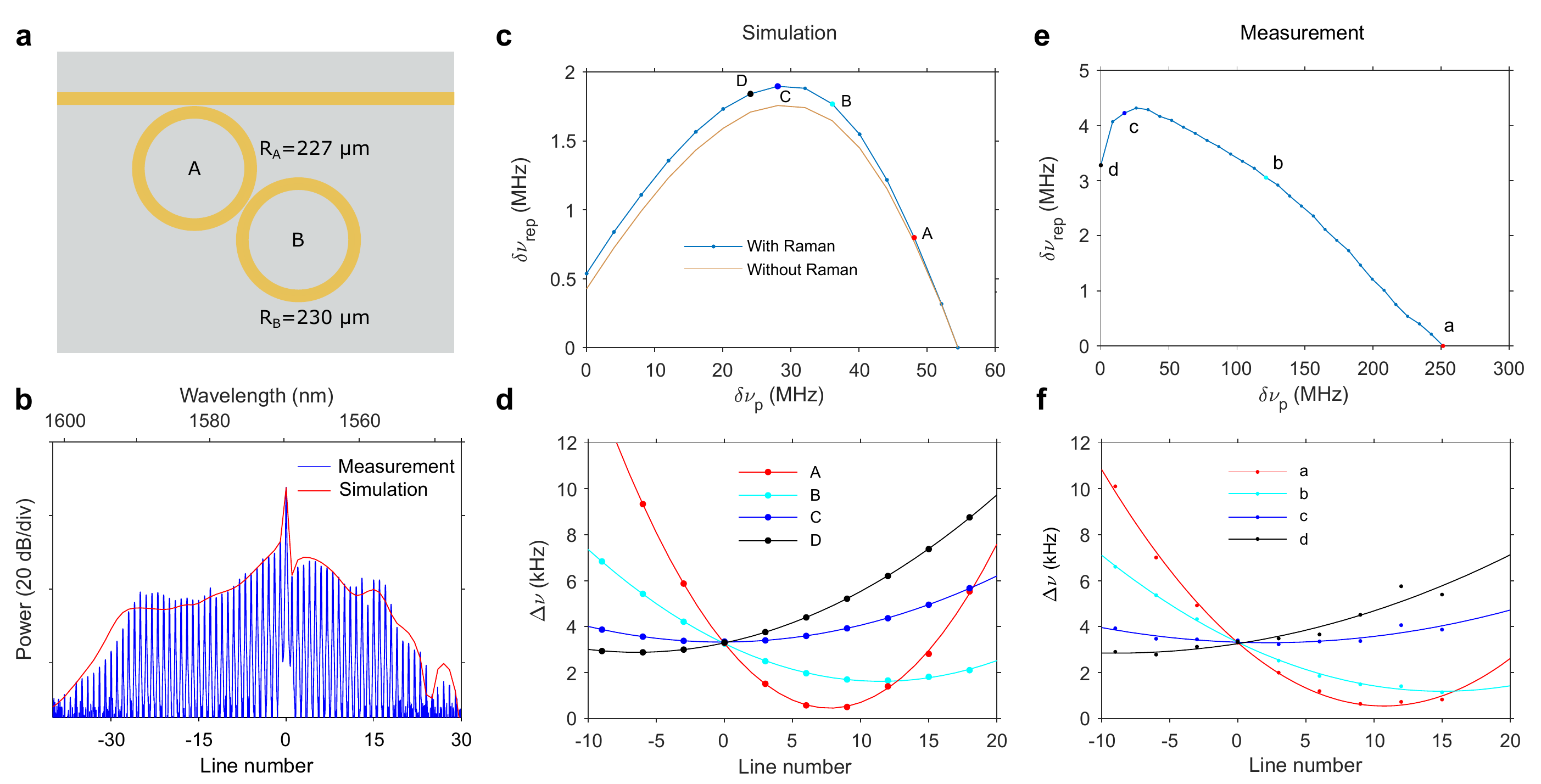}
\caption{{Lorentzian linewidth of photonic molecule soliton microcombs.} \textbf{a} Schematic of photonic molecule microresonators.  \textbf{b} Optical spectrum of the 
photonic molecule microcomb. \textbf{c}  Simulated dependence of soliton microcomb repetition rate change on the relative pump frequency. \textbf{d} Simulated Lorentzian linewidth of comb lines at different pump frequencies. \textbf{e}-\textbf{f} are measured results corresponding to \textbf{c} and \textbf{d}, respectively.}
\label{fig:s1}
\end{figure*}

In contrast to the single-cavity microcomb in the main text, the  repetition rate does not follow a linear but a parabolic dependence of the pump frequency, see Supplementary Fig. 2c. It means there exists a quiet point in the system. Note that this quiet point is mainly induced by the balance of dispersive waves, as the Raman plays a quite weak effect here (with the Raman term switched off in the simulation, we can still obtain this quiet point, see Supplementary Fig. 2c).  The details of the  underlying physics  deserves further exploration.
Here we focus on its impact on the Lorentzian linewidth of comb lines.  At the quiet point, there is no significant linewidth narrowing effect as the linewidth of comb lines turns out to be slightly higher than that of the pump due to the shot noise and pump intensity noise. As shown in Supplementary Fig. 2d, we observe the linewidth of comb lines is also a function of the detuning, and the number of the comb line with the minimum linewidth can be modified to be at either the blue or the red side of pump.


These simulated phenomena are also qualitatively confirmed in the experiment, as seen in Supplementary Fig. 2e and 2f. The difference between the simulations and experiments can be attributed to the thermal effect which is not included in the simulations. Unlike the generation of soliton in a single  microresonator with anomalous dispersion where the pump is red detuned from the cavity resonance, here the pump is at the blue detuned side of the cavity resonance [5]. Therefore, the system's temperature strongly relies on the pump frequency. The change of temperature would induce the two-cavity detuning and repetition rate change, and even modify the soliton existence condition. 

The above results indicate that the elastic tape model appears to apply for a broad range of coherent microcombs, in spite of different physical mechanisms coupling with repetition rate with detuning.



\section*{\textbf{Supplementary Note 2: Influence of the pump linewidth}}

\begin{figure*}[th]
\centering
\includegraphics[width=\linewidth]{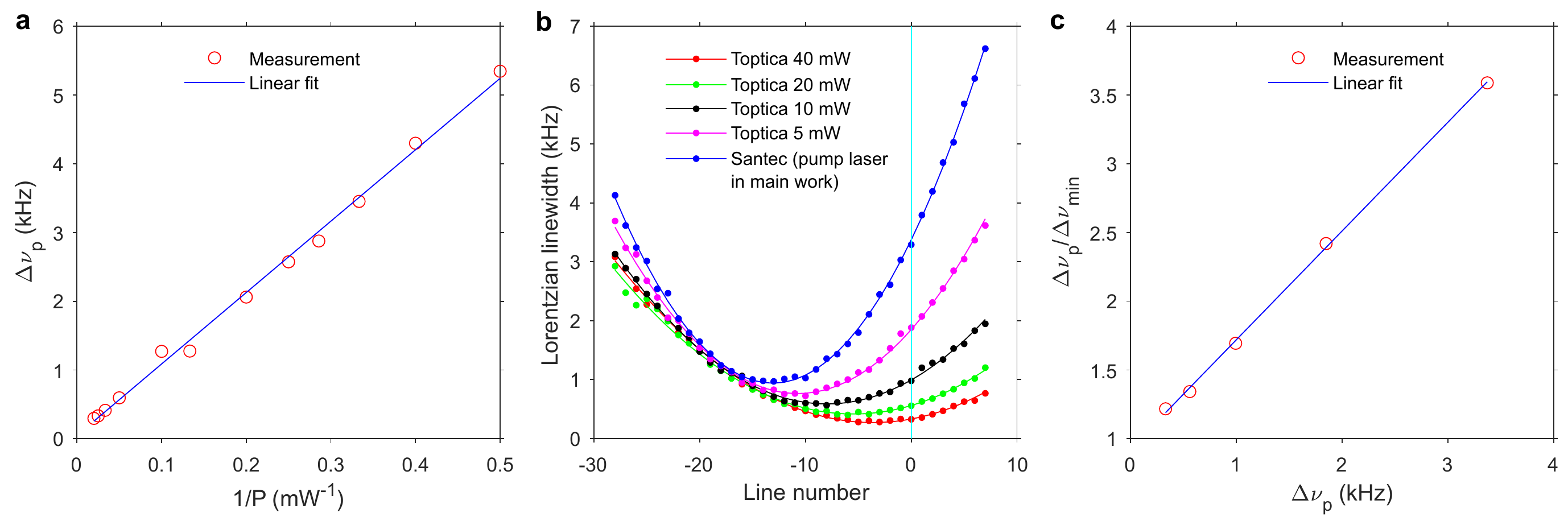}
\renewcommand{\figurename}{{Supplementary Figure}}
\caption{ Operation of soliton microcomb with a pump with different Lorentzian linewidth. \textbf{a} The measured Lorentzian linewidth of a pump laser (Toptica, CTL 1550) at different output powers. \textbf{b} The measured Lorentzian linewidth of the soliton microcomb with different pump sources. \textbf{c} Relative reduction of Lorentzian linewidth as a function of the pump Lorentzian linewidth. }
\label{fig:s2}
\end{figure*}
According to equation (9) in the main text, the Lorentzian linewidth reduction is also a function of the Lorentzian linewidth of the pump. Here, we demonstrate this phenomenon experimentally. We use a different pump laser (Toptica, CTL 1550) whose Lorentzian linewidth  decreases inversely proportional to pump power (see Supplementary Fig. 3a) [6].  

Supplementary Fig. 3b presents the Lorentzian linewidth distribution of a soliton microcomb when the microresonator is pumped with a laser featuring different Lorentzian linewidths. As the pump linewidth decreases, the position of the quiet mode shifts towards the pump location. As plotted in Supplementary Fig. 3c, the Lorentzian linewidth reduction attained at the quiet mode increases linearly with the value of the Lorentzian linewidth of the pump. These trends are in agreement with the predictions set forth by equation (9) in the main text.

\begin{figure*}[!h]
\centering
\includegraphics[width=0.9\linewidth]{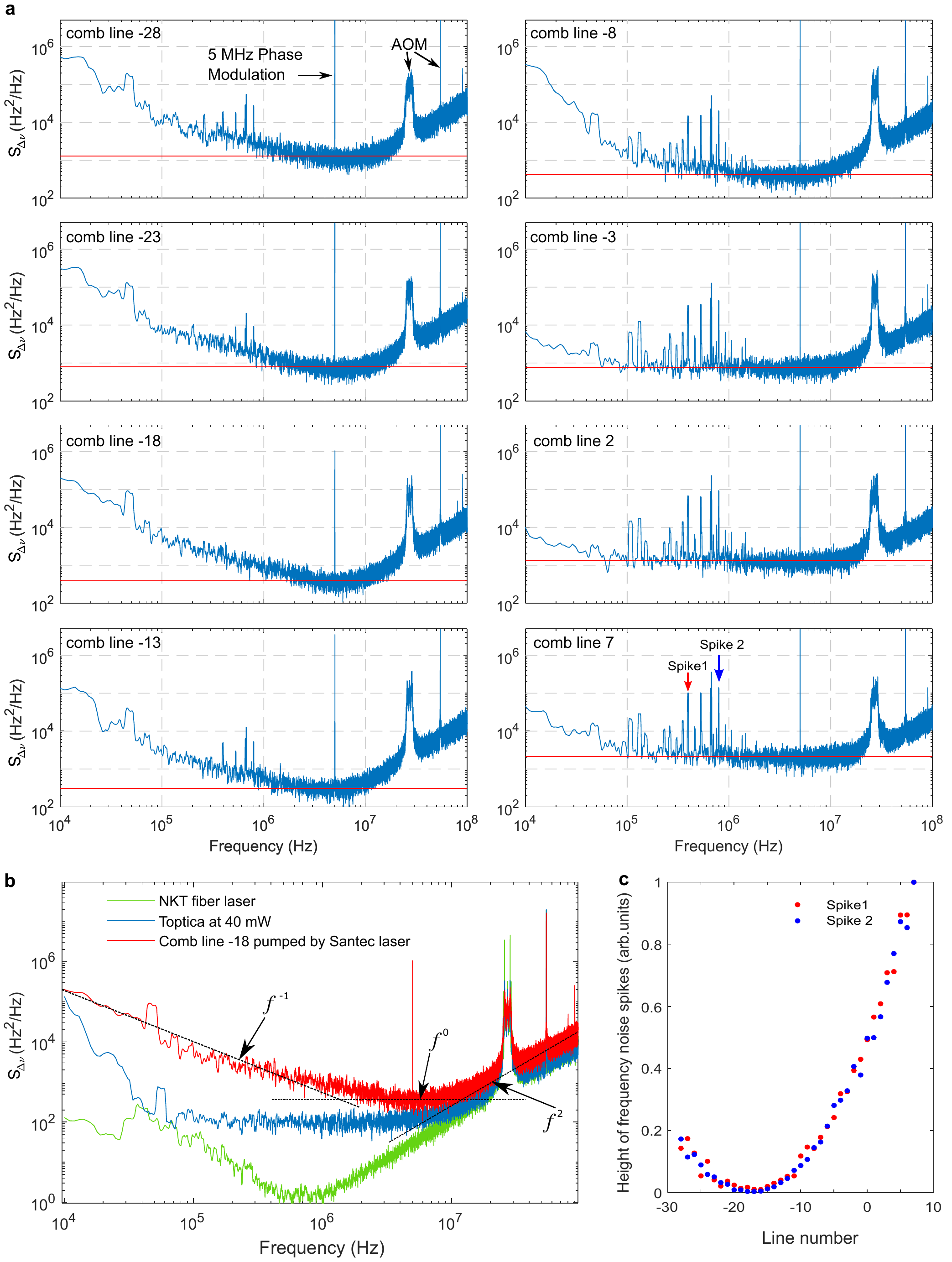}
\renewcommand{\figurename}{{Supplementary Figure}}
\caption{\textbf{a} Frequency-noise power spectral density (PSD) $S_{\Delta\nu}$ of selected microcomb lines. The Lorentzian linewidth of comb lines are calculated based on the mean value of white-noise $f^0$ plateau in the PSD, as marked with the red lines. \textbf{b} Measured frequency-noise PSD of three lasers for comparison. \textbf{c} The relative height of two frequency noise spikes originating from pump laser. Spikes 1 and 2 are indicated in \textbf{a} (comb line 7).}
\label{fig:s4}
\end{figure*}

\section*{\textbf{Supplementary Note 3: Frequency noise power spectral density and Lorentzian linewidth of the comb lines}}

The Lorentzian linewidth of a laser is usually obtained from its frequency noise power spectral density (PSD) $S_{\Delta\nu}(f)$. Here the frequency noise PSD for some representative comb lines are shown in Supplementary Fig. 4, which corresponds to the experimental results presented in Fig. 3 in the main text. 

The Lorentzian linewidths of the comb lines are obtained based on the average value of the white-noise plateau within the 3-5 MHz region of $S_{\Delta\nu}(f)$,  marked as the red lines in Supplementary Fig. 4. This region is chosen based on two considerations. Firstly, as can be seen from Supplementary Fig. 4{a}, the flicker  noise caused by thermorefractive noise and pump technical noise dominates at low frequencies. Secondly, at high frequencies, the measured $S_{\Delta\nu}(f)$ is dominated by the white phase noise, which originates from the optical amplifiers ASE noise and the thermal noise of the measurement system [7].  The white phase noise corresponds to $f^2$ frequency noise, as shown in Supplementary Fig. 4{b}. To verify that the measured values are much higher than the measurement floor of our system, we measured the frequency noise PSD of other low linewidth lasers, which are shown in Supplementary Fig. 4{b}. It is shown that our measurement system can capture the frequency noise PSD $S_{\Delta\nu}(f)$ accurately up to ~5 MHz.

Besides monitoring the white-noise plateau, one can see there are some frequency noise spikes in most of comb lines, which originates from the pump laser. Similar to the experiment presented in the main text, the magnitude of the those spikes changes with line number and gets suppressed at the fixed point, see Supplementary Fig. 4{c}.

\section*{\textbf{Supplementary Note 4: Pump at another wavelength}}
\begin{figure*}[!h]
    \centering
    \includegraphics[width=\linewidth]{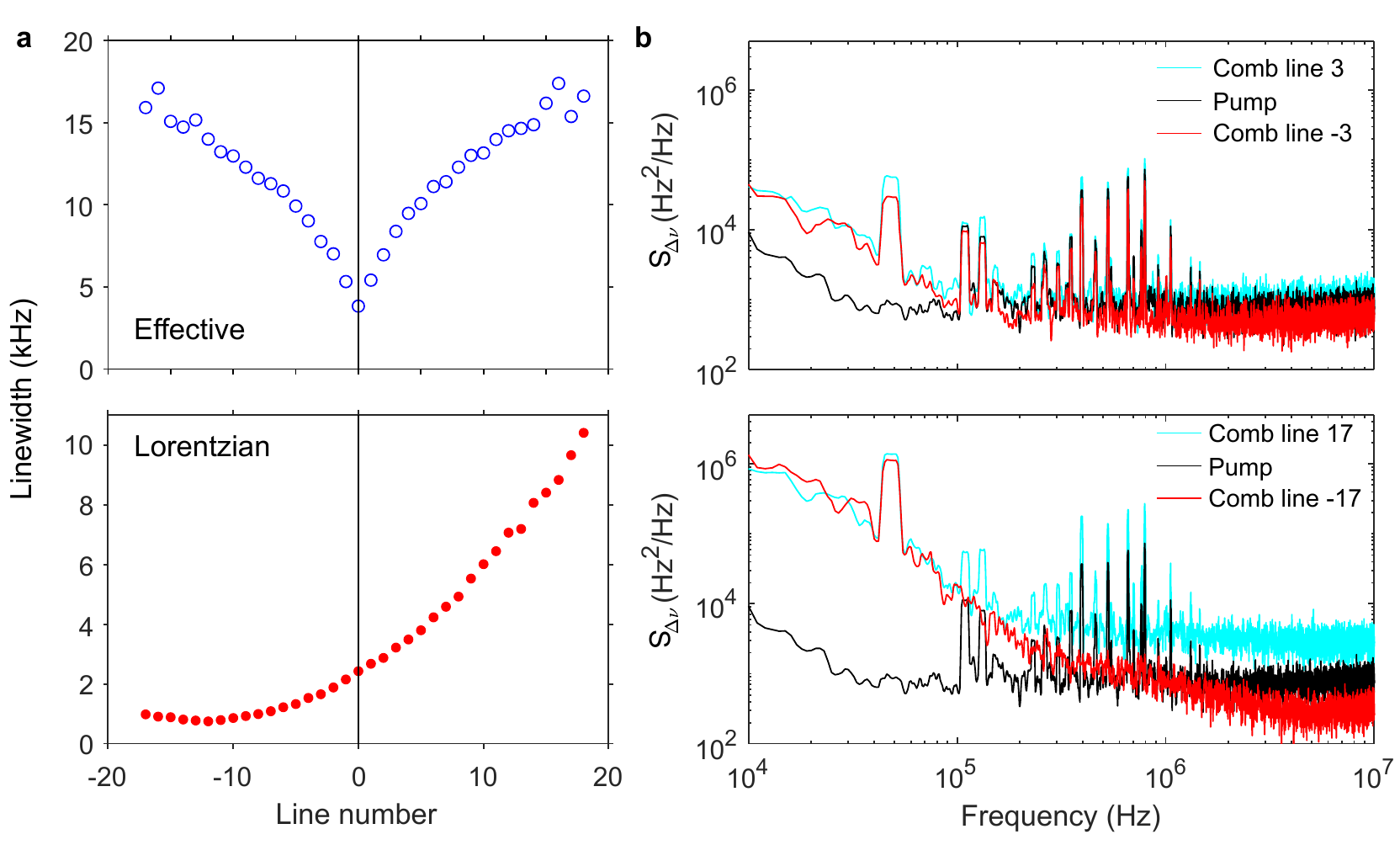}
    \renewcommand{\figurename}{{Supplementary Figure}}
    \caption{ \textbf{a} The effective linewidth and the Lorentzian linewidth when the pump (0 line number) is located at 1561 nm instead. \textbf{b} Frequency noise PSD for two pairs of comb lines symmetrically located close (top panel) and far (bottom panel) from the pump. }
    \label{fig:s4}
\end{figure*}
To perform a better comparison of the Lorentzian linewidth and effective linewidth, here we provide further experimental results  by pumping a different mode close to 1561 nm, i.e. close to the center bandwidth of our measurement system. The measured effective linewidth and Lorentzian linewidth of the comb lines are shown in Supplementary Fig. 5a. 

The symmetric distribution of the effective linewidth around the pump can be understood from the frequency noise PSD, see Supplementary Fig. 5b. Because the effective linewidth is dominated by the low-frequency region in the PSD, they are nearly the same for two comb lines with the same frequency distance from the pump, e,g., comb lines -17 and 17.

\section*{\textbf{Supplementary References}}
1. V. Brasch, M. Geiselmann, T. Herr, G. Lihachev, M.H. Pfeiffer, M.L. Gorodetsky, T.J. Kippenberg. Science \textbf{351} 357-360 (2016)\\
2. A.B. Matsko, W. Liang, A.A. Savchenkov, D. Eliyahu and L. Maleki. Opt. Lett. \textbf{41} 2907-2910 (2016)\\
3. Q.F. Yang, X. Yi, K.Y. Yang and K. Vahala,  Optica,  \textbf{3}, 1132-1135 (2016).\\
4. X. Yi, Q.F. Yang, X. Zhang, K.Y. Yang, X. Li and K. Vahala, Nat. Commun. \textbf{8}, 14869 (2017).\\
5. Ó. B. Helgason, F.R. Arteaga-Sierra, Z. Ye, K. Twayana, P. A. Andrekson, M. Karlsson, J. Schr{\"o}der, and V. Torres-Company. Nature Photonics \textbf{15} 305-310  (2021).\\
6. K. Petermann, \textit{Laser diode modulation and noise}, Vol.3 (Springer Science \& Business Media, 1991). \\
7. N. R. Newbury and W. C. Swann, J. Opt. Soc. Am. B  \textbf{24}, 1756 (2007).\\

\end{document}